\documentstyle[equations,12pt]{article}
\begin{document}

\begin{flushright}
IPNO/TH 94-85
\end{flushright}
\vspace{1 cm}
\begin{center}
{\large {\bf How to obtain a covariant Breit type equation \\
from relativistic Constraint Theory}}
\vspace{1 cm}

J. Mourad \\
{\it Laboratoire de Mod\`eles de Physique Math\'ematique, \\
Universit\'e de Tours, Parc de Grandmont, \\
F-37200 Tours, France} \\
\vspace{1 cm}
H. Sazdjian \\
{\it Division de Physique Th\'eorique\footnote{Unit\'e de Recherche des
Universit\'es Paris 11 et Paris 6 associ\'ee au CNRS.},
Institut de Physique Nucl\'eaire, \\
Universit\'e Paris XI, \\
F-91406 Orsay Cedex, France}
\end{center}
\vspace{8 cm}

\noindent
October 1994
\newpage

\begin{center}
{\large Abstract}
\end{center}
It is shown that, by an appropriate modification of the structure of
the interaction potential, the Breit equation can be incorporated into
a set of two compatible manifestly covariant wave equations, derived
from the general rules of Constraint Theory. The complementary
equation to the covariant Breit type equation determines the evolution
law in the relative time variable. The interaction potential can be
systematically calculated in perturbation theory from Feynman diagrams.
The normalization condition of the Breit wave function is determined.
The wave equation is reduced, for general classes of potential, to a
single Pauli-Schr\"odinger type equation. As an application of the
covariant Breit type equation, we exhibit massless pseudoscalar bound state
solutions, corresponding to a particular class of confining potentials.

\newpage

\section{Introduction}
Historically, The Breit equation \cite{b} represents the first attempt
to describe the relativistic dynamics of two interacting fermion systems.
It consists in summing the free Dirac hamiltonians of the two fermions
and adding mutual and, eventually, external potentials. This equation,
when applied to QED, with one-photon exchange diagram considered in
the Coulomb gauge, and solved in the linearized approximation, provides
the correct spectra to order $\alpha ^4$ \cite{b,bs} for various bound
state problems.\par
However, attempts to improve the predictivity of the equation, by
developing from it a systematic perturbation theory or by solving it
exactly, have failed, due to its inability to incorporate the whole
effects of the interaction hamiltonian of QED \cite{bs}. Also, this
equation does not satisfy global charge conjugation symmetry \cite{s}.
Finally, the Breit equation, as it stands, is not relativistically
invariant, although one might consider it valid in the c.m. frame.\par
Nevertheless, despite these drawbacks, the Breit equation has remained
popular. The main reason for this is due to the fact that it is a
differential equation in $x$-space and thus it permits the study of
effective local potentials with standard techniques of quantum mechanics.
Improvements of this equation usually transform it into integral
equations in momentum space, because of the presence of projection
operators \cite{s,mw}.\par
The purpose of this article is to derive, from relativistic Constraint
Theory \cite{ll}, a covariant Breit type equation, where the free part is the
sum of individual Dirac hamiltonians. The latter framework ensures the
relativistic invariance of the equations describing two-particle
systems with mutual interactions \cite{cva1,s1} and establishes the
connection with Quantum Field Theory and the Bethe-Salpeter equation
by means of a Lippmann-Schwinger-Quasipotential type equation relating
the potential to the off-mass shell scattering amplitude \cite{s2}.\par
The fact that a covariant Breit type equation can be obtained from
Constraint Theory was already shown by Crater and Van Alstine \cite{cva2}.
In the present paper we show that, for general classes of interaction, the
covariant Breit type equation is {\it  equivalent} to the Constarint
Theory wave equations, provided it is supplemented with a second
equation which explicitly eliminates the relative energy variable and
at the same time ensures Poincar\'e invariance of the theory. The
potential that appears in the main equation has a c.m. energy dependence
that also ensures the global charge conjugation symmetry of the system.\par
The paper is organized as follows. In Sec. 2, we derive the covariant Breit
type equation from the relativistic Constraint Theory wave equations.
In Sec. 3, we determine the normalization condition of the wave function.
In Sec. 4, we reduce the wave equation, relative to a sixteen-component
wave function, to a Pauli-Schr\"odinger type equation, relative to a
four-component wave function. As an application of the covariant Breit
equation, we exhibit, in Sec. 5, massless pseudoscalar bound state
solutions, corresponding to a particular class of confining potentials,
involving mainly pseudoscalar and spacelike vector interactions.
Conclusion follows in Sec. 6.\par

\section{The covariant Breit equation}
\setcounter{equation}{0}

We begin with the Constraint Theory wave equations describing a
system of two spin-$1/2$ particles composed of a fermion of mass $m_1$
and an antifermion of mass $m_2\ $, in mutual interaction \cite{s1} :
\subequations
\begin{eqnarray}
\label{2e1a}
\big (\ \gamma _1.p_1 - m_1\big )\ \widetilde \Psi \ =\
\big (-\gamma _2.p_2 + m_2\big )\ \widetilde V\ \widetilde \Psi \ ,\\
\label{2e1b}
\big (-\gamma _2.p_2 - m_2\big )\ \widetilde \Psi \ =\
\big (\ \gamma _1.p_1 + m_1\big )\ \widetilde V\ \widetilde \Psi \ .
\end{eqnarray}
\endsubequations
Here, $\widetilde \Psi$ is a sixteen-component spinor wave function of rank
two and is represented as a $4\times 4$ matrix :
\begin{equation} \label{2e2}
\widetilde \Psi \ =\ \widetilde \Psi _{\alpha _1 \alpha _2} (x_1,x_2)\ \ \ \
(\alpha _1,\alpha _2=1,\ldots ,4)\ ,
\end{equation}
where $\alpha _1 (\alpha _2)$ refers to the spinor index of particle 1(2).
$\gamma _1$ is the Dirac matrix $\gamma $ acting in the subspace of the
spinor of particle 1 (index $\alpha _1$); it acts on $\widetilde \Psi $
from the left. $\gamma _2$ is the Dirac matrix acting in the subspace of
the spinor of particle 2 (index $\alpha _2$); it acts on $\widetilde \Psi $
from the right; this is also the case of products of $\gamma _2$ matrices,
which act on $\widetilde \Psi$ from the right in the reverse order :
\begin{eqnarray}   \label{2e3}
\gamma _{1\mu} \widetilde \Psi & \equiv &(\gamma _\mu )_{\alpha _1 \beta _1}
\widetilde \Psi _{\beta _1 \alpha _2}\ ,\ \ \
\gamma _{2\mu} \widetilde \Psi \equiv \widetilde \Psi _{\alpha _1 \beta _2}
(\gamma _\mu )_{\beta _2 \alpha _2}\ ,\nonumber \\
\gamma _{2\mu}\gamma _{2\nu} \widetilde \Psi & \equiv &
\widetilde \Psi _{\alpha _1 \beta _2} (\gamma _\nu \gamma _\mu)_
{\beta _2 \alpha _2}\ ,\ \ \
\sigma _{a\alpha \beta} = \frac {1}{2i} \big [ \gamma _{a\alpha },
\gamma _{a\beta } \big ]\ \ \ (a=1,2)\ .
\end{eqnarray}
In Eqs. (2.1) $p_1$ and $p_2$ represent the momentum operators of particles
1 and 2, respectively. $\widetilde V$ is a Poincar\'e invariant potential.\par
The compatibility (integrability) condition of the two equations (2.1)
imposes conditions on the wave function and the potential. For the
wave function, one finds the constraint :
\begin{equation} \label{2e4}
\big [\ (p_1^2 - p_2^2) - (m_1^2 - m_2^2)\ \big ]\ \widetilde \Psi \ =\ 0\ ,
\end{equation}
which allows one to eliminate the relative energy variable in a covariant
form. For eigenfunctions of the total momentum operator $P$, the solution
of Eq. (\ref{2e4}) is :
\begin{equation} \label{2e5}
\widetilde \Psi \ =\ e^{\displaystyle -iP.X}\
e^{\displaystyle -i(m_1^2 - m_2^2) P.x/(2P^2)}\
\widetilde \psi (x^T)\ ,
\end{equation}
where we have used notations from the following definitions :
\begin{eqnarray} \label{2e6}
P &=& p_1 + p_2\ ,\ \ \ p = \frac {1}{2} (p_1 - p_2)\ ,\ \ \
M = m_1 + m_2\ , \nonumber \\
X &=& \frac {1}{2} (x_1 + x_2)\ ,\ \ \ x = x_1 - x_2\ .
\end{eqnarray}
We also define transverse and longitudinal components of four-vectors with
respect to the total momentum $P$ :
\begin{eqnarray} \label{2e7}
q_\mu ^T &=& q_\mu - \frac {(q.P)}{P^2} P_\mu\ ,\ \ \
q_\mu ^L = (q.\hat P) \hat P_\mu \ ,\ \ \ \hat P_\mu = P_\mu /\sqrt {P^2}\ ,
\nonumber \\
q_L &=& q.\hat P\ ,\ \ \ P_L = \sqrt {P^2}\ .
\end{eqnarray}
This decomposition is manifestly covariant. In the c.m. frame the transverse
components reduce to the three spacelike components, while the longitudinal
component reduces to the timelike component of the corresponding four-vector.
(Note that $x^{T2} = - {\bf x}^2$ in the c.m. frame.)
Also notice that, with the definition of the longitudinal components,
$P_L$, which is the positive square root of $P^2$, does not change
sign for negative energy states (under the change $P\rightarrow -P$);
in this case, it is the longitudinal components, $q_L$, of those
four-vectors which are independent of $P$ that change sign, since these
are linear functions of $\hat P$.\par
For the potential, one finds the constraint :
\begin{equation} \label{2e8}
\big [\ p_1^2 - p_2^2\ , \widetilde V\ \big ]\ \widetilde \Psi \ =\ 0\ ,
\end{equation}
which means that $\widetilde V$ is independent of the relative
longitudinal coordinate $x_L$ :
\begin{equation} \label{2e9}
\widetilde V\ =\ \widetilde V(x^T,\ P_L,\ p^T,\ \gamma _1,\ \gamma _2)\ .
\end{equation}
\par
Equations (\ref{2e5}) and (\ref{2e9}) show that the
internal dynamics of the system is three-dimensional, besides the spin degrees
of freedom, described by the three-dimensional transverse coordinate $x^T$.
\par
The relationship between the potential $\widetilde V$ and Feynman diagrams
is summarized by the following
Lippmann-Schwinger-Quasipotential type
\cite{ltlttk,blsu,g,pl,f,fh,t,lcl,s2,ms} equation :
\begin{eqnarray} \label{2e10}
&\ &\widetilde V\ -\ \widetilde T\ -\ \widetilde V G_0 \widetilde T\ =\ 0\ ,\\
&\ &\widetilde T(P,p^T,p'^T)\ \equiv \ \frac {i}{2P_L} \bigg [\ T(P,p,p')\
\bigg ]_{C(p),C(p')}\ ,\nonumber
\end{eqnarray}
where :\\
i) $T$ is the off-mass shell fermion-antifermion scattering
amplitude;\\
ii) $C$ is the constraint (\ref{2e4}) :
\begin{equation} \label{2e11}
C(p)\ \equiv \ (p_1^2 - p_2^2) - (m_1^2 - m_2^2)\ =\ 2P_Lp_L - (m_1^2 - m_2^2)
\ \approx \ 0\ ;
\end{equation}
in Eq. (\ref{2e10}) the external momenta of the amplitude $T$ are submitted to
the constraint $C$;\\
iii) $G_0$ is defined as :
\begin{equation} \label{2e12}
G_0(p_1,p_2)\ =\ \widetilde S_1(p_1)\ \widetilde S_2(-p_2)\ H_0\ ,
\end{equation}
where $\widetilde S_1$ and $\widetilde S_2$ are the propagators of the
two fermions, respectively, in the presence of the constraint
(\ref{2e11}), and $H_0$ is the Klein-Gordon operator, also in the presence of
the constraint (\ref{2e11}) :
\begin{equation} \label{2e13}
H_0\ =\ (p_1^2 - m_1^2) \bigg \vert _C \ =\ (p_2^2 - m_2^2) \bigg \vert _C \ =\
\frac{P^2}{4} - \frac{1}{2}(m_1^2 + m_2^2) + \frac {(m_1^2 - m_2^2)^2}
{4P^2} + p^{T2}\ .
\end{equation}
\par
In order to obtain the covariant Breit equation, we define the covariant
Dirac ``hamiltonians'' :
\subequations
\begin{eqnarray}
\label{2e14a}
{\cal H}_1\ &=&\ m_1\gamma _{1L} - \gamma_{1L}\gamma_1^T.p_1^T\ ,\\
\label{2e14b}
{\cal H}_2\ &=& -m_2\gamma _{2L} - \gamma_{2L}\gamma_2^T.p_2^T\ .
\end{eqnarray}
\endsubequations
We then multiply Eq. (\ref{2e1a}) by $\gamma_{1L}$ and Eq. (\ref{2e1b})
by $\gamma_{2L}\ $, respectively. After subtracting the two equations
from each other, we obtain the equation :
\begin{equation} \label{2e15}
\left ( P_L - ({\cal H}_1 + {\cal H}_2) \right )\widetilde \Psi \ =\
-\left ( P_L + ({\cal H}_1 + {\cal H}_2) \right )
( \gamma_{1L} \gamma_{2L} \widetilde V )\widetilde \Psi \ ,
\end{equation}
which can be rewritten as :
\begin{equation} \label{2e16}
\left [\ P_L (1+\gamma_{1L}\gamma_{2L}\widetilde V)\ -\
({\cal H}_1+{\cal H}_2)(1-\gamma_{1L}\gamma_{2L}\widetilde V)\ \right ]
\widetilde \Psi \ =\ 0\ .
\end{equation}
\par
Addition of the two equations to each other leads to the equation :
\begin{equation} \label{2e17}
\left (2p_L - ({\cal H}_1 - {\cal H}_2)\right )\widetilde \Psi \ =\
\left (2p_L + ({\cal H}_1 - {\cal H}_2)\right )
(\gamma_{1L} \gamma_{2L} \widetilde V) \widetilde \Psi \ ,
\end{equation}
which can be rewritten as :
\begin{equation} \label{2e18}
\left [\ 2p_L (1-\gamma_{1L}\gamma_{2L}\widetilde V)\ -\
({\cal H}_1 - {\cal H}_2)(1+\gamma_{1L}\gamma_{2L}\widetilde V)\ \right ]
\widetilde \Psi \ =\ 0\ .
\end{equation}
\par
Upon multiplying this equation by $({\cal H}_1 + {\cal H}_2)$ and
noticing that ${\cal H}_1^2 - {\cal H}_2^2 = m_1^2 - m_2^2\ $, it becomes,
after using Eq. (\ref{2e16}) :
\begin{equation} \label{2e19}
\left [\ 2p_LP_L - (m_1^2 - m_2^2)\ \right ] (1+\gamma_{1L}\gamma_{2L}
\widetilde V) \widetilde \Psi \ =\ 0\ .
\end{equation}
\par
We now define the Breit wave function $\Psi _B$ by :
\begin{equation} \label{2e20}
\Psi_B\ =\ (1-\gamma_{1L}\gamma_{2L}\widetilde V)\widetilde \Psi \ .
\end{equation}
Then, Eq. (\ref{2e16}) takes the form :
\begin{equation} \label{2e21}
\left [\ P_L(1+\gamma_{1L}\gamma_{2L}\widetilde V)
(1-\gamma_{1L}\gamma_{2L}\widetilde V)^{-1} - ({\cal H}_1 + {\cal H}_2)
\ \right ] \Psi_B \ =\ 0\ ,
\end{equation}
while Eq. (\ref{2e19}) yields, after factorizing the term
$(1+\gamma_{1L}\gamma_{2L}\widetilde V)(1-\gamma_{1L}\gamma_{2L}
\widetilde V)^{-1}$ :
\begin{equation} \label{2e22}
\left [\ 2p_LP_L - (m_1^2 - m_2^2)\ \right ] \Psi_B\ =\ 0\ .
\end{equation}
\par
Equations (\ref{2e21}) and (\ref{2e22}) are the two wave equations
satisfied by the Breit wave function $\Psi_B\ $. As far as the wave
function transformation (\ref{2e20}) is nonsingular, they are
{\it equivalent} to the initial two wave equations (2.1) of Constraint
Theory.\par
Equation (\ref{2e21}) is the obvious generalization of the Breit
equation. Its interaction dependent part has an explicit c.m. energy
$(P_L)$ dependence which restores the global charge conjugation symmetry
that was lacking in the Breit equation. For each solution of Eq. (\ref{2e21})
with total momentum $P\ $, there will correspond, for charge conjugation
invariant interactions, a charge conjugated solution with momentum $-P\ $.
\par
Equation (\ref{2e22}) determines the relative time evolution law of the
wave function, as in Eq. (\ref{2e5}), and ensures the relativistic
invariance of the theory. While Eq. (\ref{2e21}) might be
considered alone in the c.m. frame, Eq. (\ref{2e22}) indicates the way
of passing to other reference frames.\par
In the c.m. frame, with the standard definitions $\beta =\gamma _0$ and
\mbox{\boldmath $\alpha $}=$\gamma_0$\mbox{\boldmath $\gamma $},
Eq. (\ref{2e21}) becomes :
\begin{equation} \label{2e23}
\left [\ P_0 (1+\beta_1\beta_2\widetilde V)
(1-\beta_1\beta_2\widetilde V)^{-1} - (m_1\beta_1 +
\mbox{\boldmath $\alpha_1 .p$}
- m_2\beta_2 - \mbox{\boldmath $\alpha_2 .p$})\ \right ]\Psi_B\ =\ 0\ .
\end{equation}
\par
In perturbation theory, $\widetilde V$ has, in lowest order, according
to Eq. (\ref{2e10}), the structure \cite{ms} :
\begin{equation} \label{2e24}
\widetilde V\ =\ -\frac {1}{2\sqrt{P^2}} U(x^T,\gamma_1,\gamma_2)\ ,
\end{equation}
where $U$ is the three-dimensionally reduced form of the propagator of
the exchanged particle, including the couplings at the vertices. To this
order, Eq. (\ref{2e23}) takes the form:
\begin{equation} \label{2e25}
\left [\ P_0 - \beta_1\beta_2\epsilon (P_0) U - (m_1\beta_1
+ \mbox{\boldmath $\alpha_1 .p$} - m_2\beta_2 -
\mbox{\boldmath $\alpha_2 .p$})\ \right ]
\Psi_B \ =\ 0\ .
\end{equation}
We notice here, in distinction from the Breit equation, the presence of
the energy sign factor in front of the potential $U\ $; it is this factor
which ensures the global charge conjugation symmetry of the
equation.\par
Finally, in the limit when $m_2$ tends to infinity, Eq. (\ref{2e25})
yields the Dirac equation of particle 1, with the potential $\beta_1 U$
($\beta_2$ is replaced by $-1$ for the antifermion and $\epsilon (P_0)$
by $+1$ in this limit).\par
Equations (2.1), or equivalently (\ref{2e21}) and (\ref{2e22}), were
analyzed, in Ref. \cite{ms}, in the nonrelativistic limit, to order
$1/{c^2}\ $, in particular for the electromagnetic interaction case. For an
arbitrary covariant gauge of the photon propagator, the corresponding
hamiltonian receives contributions (among others) from quadratic terms
generated by the one-photon exchange diagram as well as from the two-photon
exchange diagrams. However, it turns out that in the Coulomb gauge (and
also in the Landau gauge to that order) the two-photon contribution cancels
the quadratic terms arising from the one-photon exchange diagram and one
then is left with the Breit hamiltonian \cite{bs,blp}. This explains why
the Breit equation in its linearized approximation provides a correct
result to order $\alpha ^4\ $. However, in other gauges than the Coulomb
and Landau gauges, it is necessary to take into account the quadratic terms
as well as the two-photon exchange contribution to obtain a correct
result.\par

\section{Normalization condition}
\setcounter{equation}{0}

The normalization condition of the wave function $\widetilde \Psi$
can be determined either from the construction
of tensor currents of rank two, satisfying two
independent conservation laws, with respect to $x_1$ and $x_2$ \cite{s1},
or from the integral equation of the corresponding
Green's function \cite{f,lcl}. One finds for the norm of
$\widetilde \psi$ [Eq. (\ref{2e5})] the formula (in the c.m. frame and
for local potentials in $x^T$) :
\begin{equation} \label{3e1}
\int d^3{\bf x}\ Tr \bigg \{ \widetilde \psi ^{\dagger }\ \big [1 - \widetilde
V^{\dagger } \widetilde V + 4\gamma _{10} \gamma _{20} P_0^2 \frac
{\partial \widetilde V}{\partial P^2} \big ]\ \widetilde \psi \bigg \}\ =\
2P_0\ ,
\end{equation}
where $\widetilde V$ satisfies the hermiticity condition :
\begin{equation} \label{3e1p}
\widetilde V^{\dagger} = \gamma_{10} \gamma_{20} \widetilde V \gamma_{10}
\gamma_{20}\ .
\end{equation}
For energy independent potentials (in the c.m. frame) the norm of
$\widetilde \psi$ is not positive definite for arbitrary $\widetilde V$.
In order to ensure positivity,
it is sufficient that the potential $\widetilde V$
satisfy the inequality
\begin{equation} \label{3e2}
\frac{1}{4} Tr (\widetilde V^{\dagger } \widetilde V) \ < \ 1\ .
\end{equation}
In this case one is allowed to make the wave function transformation
\begin{equation} \label{3e3}
\widetilde \Psi\ =\ \big [ 1 - \widetilde V^{\dagger } \widetilde V \big ]
^{-\frac {1}{2}}\ \Psi
\end{equation}
and to reach a representation where the norm
for c.m. energy independent potentials is the free norm.\par
In this respect, the parametrization suggested by Crater and
Van Alstine \cite{cva3}, for potentials that commute with $\gamma_{1L}
\gamma_{2L}$ (and hence $\widetilde V^{\dagger } = \widetilde V$),
\begin{equation} \label{3e4}
\widetilde V \ =\ \tanh V \ ,
\end{equation}
satisfies condition (\ref{3e2}) and allows one to bring the equations
satisfied by $\Psi$ [Eq. (\ref{3e3})] into forms analogous to the
Dirac equation, where
each particle appears as placed in the external potential created by the
other particle, the latter potential having the same tensor nature as
potential $V$ of Eq. (\ref{3e4}).\par
We shall henceforth adopt the above parametrization (\ref{3e4}). For
more general potentials that do not commute with $\gamma_{1L} \gamma_{2L}\ $,
the natural extension of parametrization (\ref{3e4}) is:
\begin{equation} \label{3e4p}
\gamma_{1L} \gamma_{2L} \widetilde V \ =\ \tanh (\gamma_{1L} \gamma_{2L} V)\ .
\end{equation}
According to Eqs. (\ref{3e1p}) and (\ref{3e3}), we shall introduce
the wave function transformation :
\begin{equation} \label{3e5}
\widetilde \Psi \ =\ \cosh (\gamma_{1L} \gamma_{2L} V)\ \Psi\ .
\end{equation}
\par
The norm of the new wave function
$\Psi$ then becomes (in the c.m. frame) :
\begin{equation} \label{3e6}
\int d^3{\bf x}\ Tr \bigg \{ \psi^{\dagger }\ \big [ 1 + 2P_0^2
\big (e^{\displaystyle -\gamma_{10} \gamma_{20} V}
\frac {\partial \ }{\partial P^2} e^{\displaystyle \gamma_{10} \gamma_{20} V}
- e^{\displaystyle \gamma_{10} \gamma_{20} V} \frac {\partial \ }{\partial
P^2} e^{\displaystyle -\gamma_{10} \gamma_{20} V} \big ) \big ]
\psi \bigg \}\ =\ 2P_0\ .
\end{equation}
(The relationship between $\Psi$ and $\psi$ is the same as in Eq.
(\ref{2e5}).)\par
Equations (2.1) then take the form :
\subequations
\begin{eqnarray}
\label{3e7a}
(\ \gamma _1.p_1 - m_1)\ \cosh (\gamma_{1L} \gamma_{2L} V) \ \Psi \ =\
(-\gamma _2.p_2 + m_2) \gamma_{1L} \gamma_{2L}
\ \sinh (\gamma_{1L} \gamma_{2L} V) \ \Psi \ ,\\
\label{3e7b}
(-\gamma _2.p_2 - m_2)\ \cosh (\gamma_{1L} \gamma_{2L} V) \ \Psi \ =\
(\ \gamma _1.p_1 + m_1) \gamma_{1L} \gamma_{2L}
\ \sinh (\gamma_{1L} \gamma_{2L} V) \ \Psi \ .
\end{eqnarray}
\endsubequations
\par
In order to determine
the normalization condition of the Breit wave function $\Psi_B\ $, we
first define from $V$ a potential $V_B$ as :
\begin{equation} \label{3e8}
V_B\ =\ \gamma_{1L}\gamma_{2L}V\ .
\end{equation}
(Notice that because of Eq. (\ref{3e1p}) $V_B^{\dagger} = V_B$ in the
c.m. frame.)
With this potential, the relationship (\ref{2e20}) takes the form :
\begin{equation} \label{3e9}
\Psi_B\ =\ \frac {e^{\displaystyle -V_B}}{\cosh V_B} \widetilde \Psi \ ,
\end{equation}
while the relationship between $\Psi_B$ and $\Psi$ [Eq. (\ref{3e5})] is :
\begin{equation} \label{3e10}
\Psi_B \ =\ e^{\displaystyle -V_B} \Psi \ .
\end{equation}
\par
The Breit type equation (\ref{2e21}) becomes :
\begin{equation} \label{3e11}
\left [\ P_L e^{\displaystyle 2V_B} - ({\cal H}_1 + {\cal H}_2)\ \right ]
\Psi_B \ =\ 0\ .
\end{equation}
\par
The normalization condition of $\psi_B$ (defined from $\Psi_B$ as in
Eq. (\ref{2e5})) is, in the c.m. frame :
\begin{equation} \label{3e12}
\int d^3 {\bf x}\ Tr \bigg \{ \psi_B^{\dagger} \big [\ e^{\displaystyle 2V_B}
+ 2P_0^2 \frac {\partial \ }{\partial P^2} e^{\displaystyle 2V_B}\ \big ]
\psi_B \bigg \} \ =\ 2P_0\ .
\end{equation}
\par
We therefore end up with three different representations for the
two-particle wave function. The first one, $\widetilde \Psi $
[Eqs. (2.1)], corresponds to the framework where Constraint Theory
conditions, as well as connection with Quantum Field Theory and the
Bethe-Salpeter equation, are most easily established. The second one,
$\Psi $ [Eqs. (\ref{3e5}) and (3.9)], corresponds to the
``canonical'' representation, for which the norm, for c.m. energy
independent potentials, is the free one. This representation is also
the one used by Crater and Van Alstine \cite{cva3}. The third one,
$\Psi_B$ [Eqs. (\ref{3e9}), (\ref{3e10}), (\ref{3e11}) and (\ref{2e22})],
corresponds to the Breit representation.\par

\section{Resolution of the Breit type equation}
\setcounter{equation}{0}

By decomposing the wave function $\Psi$ along $2\times 2$ matrix
components, Eqs. (3.9) can be solved with respect to one of these
components and transformed, for the case of potentials commuting with
$\gamma_{1L} \gamma_{2L}$, into a second order differential equation
of the Pauli-Schr\"odinger type \cite{ms}. A similar reduction is also
possible starting from the Breit type equation (\ref{3e11}).\par
The relative time dependence of the wave function being determined
by Eq. (\ref{2e22}), with a solution of the form (\ref{2e5}),
one decomposes the internal
$4\times 4$ matrix wave function $\psi_B$ on the basis of the matrices
$1,\ \gamma _L,\ \gamma _5$ and $\gamma _L \gamma _5$ by defining $2\times 2$
matrix components :
\begin{equation} \label{4e1}
\psi_B \ =\ \psi _{B1} + \gamma _L \psi _{B2} + \gamma _5 \psi _{B3} +
\gamma _L \gamma _5 \psi _{B4} \ \equiv \ \sum _{i=1}^{4} \
\Gamma _i \psi _{Bi} \ .
\end{equation}
\par
We consider the case of potentials that are local in $x^T$ (but having
eventually a c.m. energy dependence) and that are
functions of products of $\gamma _1$ and
$\gamma _2$ matrices in equal number (general vertex
corrections do not satisfy the latter property); then $V$ commutes with
$\gamma_{1L} \gamma_{2L}$ :
\begin{equation} \label{4e1p}
\gamma_{1L} \gamma_{2L} V\ =\ V \gamma_{1L} \gamma_{2L}\ .
\end{equation}
We introduce projection matrices for the above $2\times 2$ component
subspaces :
\begin{eqnarray} \label{4e2}
{\cal P}_1\ &=&\ \frac {1}{4} (1 + \gamma _{1L} \gamma _{2L})\
(1 + \gamma _{15} \gamma _{25})\ ,\ \ \ {\cal P}_2\ =\ \frac {1}{4}
(1 + \gamma _{1L} \gamma _{2L})\ (1 - \gamma _{15} \gamma _{25})\ ,\nonumber
\\
{\cal P}_3\ &=&\ \frac {1}{4} (1 - \gamma _{1L} \gamma _{2L})\
(1 + \gamma _{15} \gamma _{25})\ ,\ \ \ {\cal P}_4\ =\ \frac {1}{4}
(1 - \gamma _{1L} \gamma _{2L})\ (1 - \gamma _{15} \gamma _{25})\ .\nonumber
\\
\
\end{eqnarray}
They satisfy the relations :
\begin{equation} \label{4e3}
{\cal P}_i {\cal P}_j\ =\ \delta _{ij} {\cal P}_j \ ,\ \ \
{\cal P}_i \Gamma _j\ =\
\delta _{ij} \Gamma _j \ \ \ \ (i,j=1,\ldots ,4)\ .
\end{equation}
(The $\Gamma $'s are defined in Eq. (\ref{4e1}).)\par
Then, the most general (parity and time reversal invariant) potential
$V$ [Eqs. (\ref{3e4}) and (\ref{3e8})] we may consider has the
decomposition on the basis (\ref{4e2}) :
\begin{equation} \label{4e4}
V\ =\ \sum _{i=1}^4 a_i {\cal P}_i\ .
\end{equation}
The potentials $a_i$ themselves may still have spin dependences.
The spin operators, which act in the $2\times 2$ component subspaces, are
defined by means of the Pauli-Lubanski operators:
\begin{eqnarray} \label{4e5}
W_{1S\alpha }\ &=&\ -\frac {\hbar }{4} \epsilon _{\alpha \beta \mu \nu }
P^{\beta } \sigma _1^{\mu \nu }\ ,\ \
W_{2S\alpha }\ =\ -\frac {\hbar}{4} \epsilon _{\alpha \beta \mu \nu }
P^{\beta } \sigma _2^{\mu \nu }\ \ \ (\epsilon _{0123} = +1)\ ,\nonumber \\
W_{1S}^2 \ &=&\ W_{2S}^2 \ =\ -\frac {3}{4} \hbar ^2 P^2\ ,\ \ \
W_S\ =\ W_{1S} + W_{2S}\ .
\end{eqnarray}
They also satisfy the relations :
\begin{equation} \label{4e6}
\gamma _{1L} W_{1S\alpha }\ =\ \frac {\hbar P_L}{2} \gamma _{1\alpha }^T
\gamma _{15}\ ,\ \ \ \gamma _{2L} W_{2S\alpha }\ =\ \frac {\hbar P_L}{2}
\gamma _{2\alpha }^T \gamma _{25}\ .
\end{equation}
\par
We introduce the operators :
\begin{equation} \label{4e7}
w\ =\ (\frac {2}{\hbar P_L})^2\ W_{1S}.W_{2S}\ ,\ \ \
w_{12}\ =\ (\frac {2}{\hbar P_L})^2\ \frac {W_{1S}.x^T W_{2S}.x^T}{x^{T2}}\ ;
\end{equation}
then, the potentials $a_i$ [Eq. (\ref{4e4})] can be decomposed as :
\begin{equation} \label{4e8}
a_i\ =\ A_i\ +\ wB_i\ +\ w_{12}C_i\ \ \ \ \ (i=1,\ldots ,4)\ ,
\end{equation}
where the potentials $A_i,\ B_i,\ C_i$ are functions of $x^{T2}$
and eventually of $P^2\ $.\par
The projectors (\ref{4e2})-(\ref{4e4}) satisfy the simple property :
\begin{equation} \label{4e9}
\exp \big (\sum _{i=1}^4 a_i {\cal P}_i \big )\ =\
\sum _{i=1}^4 {\cal P}_i\ e^{\displaystyle a_i}\ .
\end{equation}
\par
The Breit potential $V_B$ [Eq. (\ref{3e8})] has also a decomposition
like (\ref{4e4}) :
\begin{equation} \label{4e10}
V_B\ =\ \sum_{i=1}^4 a_{Bi} {\cal P}_i\ ,
\end{equation}with the following relations with the $a_i$'s :
\begin{equation} \label{4e11}
a_{B1} = a_1\ ,\ \ a_{B2} = a_2\ ,\ \ a_{B3} = -a_3\ ,\ \
a_{B4} = -a_4\ .
\end{equation}
\par
The relationship (\ref{3e10}) between $\psi_B$ and $\psi$ can
be rewritten for their $2\times 2$ components as well :
\begin{equation} \label{4e12}
\psi_{Bi}\ =\ e^{\displaystyle -a_{Bi}}\ \psi_i\ \ \ \ (i=1,\ldots ,4)\ .
\end{equation}
[$\psi_i$ is defined from a decomposition of $\psi$ as in Eq. (\ref{4e1}).]
\par
The Breit type equation (\ref{3e11}) is now easily decomposed into
four equations for the four components $\psi _{Bi} \ (i=1,\ldots ,4)$ :
\subequations
\begin{eqnarray}
\label{4e13a}
P_L\ e^{\displaystyle 2a_1}\ \psi _{B1}\ &-&\
(m_1 - m_2)\ \psi _{B2}\ +\
\frac {2}{\hbar P_L}(W_{1S} - W_{2_S}).p\ \psi _{B3}\ =
\ 0\ ,\\
\label{4e13b}
P_L\ e^{\displaystyle 2a_2}\ \psi _{B2}\ &-&\
(m_1 - m_2)\ \psi _{B1}\ -\
\frac {2}{\hbar P_L}W_S.p\ \psi _{B4}\ =\ 0\ ,\\
\label{4e13c}
P_L\ e^{\displaystyle -2a_3}\ \psi _{B3}\ &-&\ M\ \psi _{B4}\
+\ \frac {2}{\hbar P_L}(W_{1S} - W_{2S}).p\ \psi _{B1}\
=\ 0\ ,\\
\label{4e13d}
P_L\ e^{\displaystyle -2a_4}\ \psi _{B4}\ &-&\ M\ \psi _{B3}\
-\ \frac {2}{\hbar P_L}W_S.p\ \psi _{B2}\ =\ 0\ .
\end{eqnarray}
\endsubequations
\par
These equations allow one to eliminate the components $\psi_{B1},\
\psi_{B2}$ and $\psi_{B4}$ in terms of $\psi_{B3}\ $, which is a
surviving component in the nonrelativistic limit. Upon defining
\begin{equation} \label{4e14}
e^{\displaystyle 2h}\ =\ 1\ -\ \frac {(m_1^2 - m_2^2)^2}{M^2 P^2}\
e^{\displaystyle -2(a_1 + a_2)} \ ,
\end{equation}
one finds for $\psi_{B1}$ and $\psi_{B2}$ the relations :
\begin{eqnarray}
\label{4e15}
P_L \psi _{B1}\ =\ e^{\displaystyle -2(a_1 + a_2 + h)}\
\big (\frac {2}{\hbar P_L}\big )\
\big \{ &-& e^{\displaystyle 2a_2}\ (W_{1S} - W_{2S}).p\
\nonumber \\
&+& \frac {(m_1^2 - m_2^2)}{M^2}\
W_S.p\ e^{\displaystyle -2a_3} \big \}\ \psi_{B3}\ ,\\
\  \nonumber \\
\label{4e16}
M \psi _{B2}\ =\ e^{\displaystyle -2(a_1 +a_2 + h)}\
\big (\frac {2}{\hbar P_L}\big )\
\big \{ &+& e^{\displaystyle 2a_1}\ W_S.p\
e^{\displaystyle -2a_3}\ \nonumber \\
&-& \frac {(m_1^2 - m_2^2)}{P^2}\
(W_{1S} - W_{2S}).p \big \}\ \psi_{B3}\ .
\end{eqnarray}
\par
One then obtains two independent equations for $\psi _{B3}$ and $\psi _{B4}$ :
\subequations
\begin{eqnarray}
\label{4e17a}
M P_L\ e^{\displaystyle -2a_4}\ \psi _{B4}\ =\ &M^2& \psi _{B3} \
+\ \big (\frac {2}{\hbar P_L}\big )^2 W_S.p\
e^{\displaystyle -2(a_1 + a_2 + h)}\nonumber \\
&\times&\big \{ +
e^{\displaystyle 2a_1}\ W_S.p\ e^{\displaystyle -2a_3}
\nonumber \\
& &\ \ -\ \frac {(m_1^2 - m_2^2)}{P^2}\
(W_{1S} - W_{2S}).p\big \}\ \psi_{B3}\ ,\\
\ & & \nonumber \\
\label{4e17b}
M P_L\ \psi _{B4}\ =\ &P^2& e^{\displaystyle -2a_3}\
\psi _{B3}\
+\ \big (\frac {2}{\hbar P_L}\big )^2 (W_{1S} - W_{2S}).p\
e^{\displaystyle -2(a_1 + a_2 + h)}\nonumber \\
&\times&\big \{ - e^{\displaystyle 2a_2}\ (W_{1S} - W_{2S}).p
\nonumber \\
& &\ \ +\ \frac {(m_1^2 - m_2^2)}{M^2}\
W_S.p\ e^{\displaystyle -2a_3}\big \}\ \psi_{B3}\ .
\end{eqnarray}
\endsubequations
\par
Elimination of $\psi _{B4}$ leads to the eigenvalue equation for
$\psi _{B3}$ :
\begin{eqnarray} \label{4e18}
\big [ P^2 e^{\displaystyle -2(a_3 + a_4)} &-& M^2\big ]\
\psi _{B3}\nonumber \\
-\big (\frac {2}{\hbar P_L}\big )^2 &W_S.p&
e^{\displaystyle -2(a_1 + a_2 + h)}\ \big \{\ e^{\displaystyle 2a_1}\ W_S.p\
e^{\displaystyle -2a_3} \nonumber \\
&-& \frac {(m_1^2 - m_2^2)}{P^2}\
(W_{1S} - W_{2S}).p\ \big \}\ \psi _{B3} \nonumber \\
-\big (\frac {2}{\hbar P_L}\big )^2 &e^{\displaystyle -2a_4}&\
(W_{1S} - W_{2S}).p\ e^{\displaystyle -2(a_1 + a_2 + h)}\
\big \{\ e^{\displaystyle 2a_2} (W_{1S} - W_{2S}).p \nonumber \\
&-& \frac {(m_1^2 - m_2^2)}{M^2}
\ W_S.p\ e^{\displaystyle -2a_3}\ \big \}\ \psi _{B3}\ =\ 0\ .
\end{eqnarray}
\par
Equation (\ref{4e18}) is a second order differential equation for the
component $\psi_{B3}\ $. Usually, by wave function transformations
one can simplify  the structure of the differential operators in it.
For the general potential (\ref{4e4}), the second order differential
operator will still exhibit a spin dependence. However, for simpler
types of potential, the spin dependence of the second order differential
operator also disappears. This is the case of the potential composed
of general combinations of scalar, pseudoscalar and vector potentials.
It has the following structure :
\begin{equation} \label{4e19}
V\ =\ V_1 + \gamma _{15} \gamma _{25} V_3 + \gamma _1^{\mu} \gamma _2^{\nu}
\ \big (\ g_{\mu \nu }^{LL} V_2 + g_{\mu \nu }^{TT} U_4 + \frac {x_{\mu }^T
x_{\nu }^T}{x^{T2}} T_4\ \big )\ ,
\end{equation}
and the decomposition of the potentials $a_i$ [Eq. (\ref{4e4})] along
these potentials is given by the relations :
\begin{eqnarray} \label{4e20}
a_1\ &=&\ V_1\ +\ V_2\ +\ V_3\ +\ wU_4\ +\ w_{12}T_4\ ,\nonumber \\
a_2\ &=&\ V_1\ +\ V_2\ -\ V_3\ -\ wU_4\ -\ w_{12}T_4\ ,\nonumber \\
a_3\ &=&\ V_1\ -\ V_2\ +\ V_3\ -\ wU_4\ -\ w_{12}T_4\ ,\nonumber \\
a_4\ &=&\ V_1\ -\ V_2\ -\ V_3\ +\ wU_4\ +\ w_{12}T_4\ .
\end{eqnarray}
\par
With potentials of the type (\ref{4e19}) and after using the wave
function transformation
\begin{eqnarray} \label{4e21}
\psi_{B3}\ =& & e^{\displaystyle V_1-V_2+V_3+2U_4+2T_4+h} \nonumber \\
&\times& \big \{-(2+\frac{W_S^2}{\hbar ^2 P^2})\
e^{\displaystyle 2V_2-U_4-T_4}\ \sinh(2U_4) \nonumber \\
& &\ \  +\frac{1}{2}(1-w_{12}) e^{\displaystyle 2V_1+U_4+T_4} +
\frac{1}{2}(1+w_{12}) e^{\displaystyle 2V_2+U_4-T_4}\big \} \ \phi_3\
\end{eqnarray}
($w_{12}$ defined in Eq. (\ref{4e7})), Eq. (\ref{4e18}) reduces to a
Pauli-Schr\"odinger type equation, where the radial differential operators
are those of the Laplace operator. This equation, which is also
obtained from a wave function transformation in the equation satisfied
by $\psi_3$ [Eq. (\ref{4e12})], was presented in Ref. \cite{ms}.\par
Equations (4.14) could also have been solved with respect to
$\psi_{B4}$ instead of $\psi_{B3}\ $.\par

\section{Zero mass solutions}
\setcounter{equation}{0}

As a straightforward application of the covariant Breit equation, with
the class of potentials considered in Sec. 4, we
shall exhibit, in this section, a class of solutions which correspond to
massless pseudoscalar bound states in the limit when the masses of the
constituent particles tend to zero.\par
The key observation is that, because of the presence of the kernel
$e^{\displaystyle 2V_B}$ in the normalization condition (\ref{3e12}),
one is allowed to search for solutions in which some of the components
$\psi _{Bi}$ are constants, provided the kernel $e^{\displaystyle 2V_B}$
is rapidly decreasing at infinity.\par
The quantum numbers of the state are determined by those of the
components $\psi_{B3}$ and $\psi_{B4}\ $, which are the surviving
components in the nonrelativistic limit. For the ground state they have
the quantum numbers $s=0$ (for the total spin operator defined in Eq.
(\ref{4e5})), $\ell =0$ (for the orbital angular momentum operator) and
$j=0$ (for the total angular momentum operator); these quantum numbers
are those of a pseudoscalar state. We shall restrict the search by
demanding that the components $\psi_{B1}$ and $\psi_{B2}$ be zero
for the ground state solution.\par
Inspection of Eqs. (\ref{4e13a}) and (\ref{4e13b}) shows that $\psi_{B3}$
must be a constant :
\begin{equation} \label{5e1}
\psi_{B3}\ =\ \phi_0\ = const.\ .
\end{equation}
(The vanishing of the components $\psi_{B1}$ and $\psi_{B2}$ can then
also be checked directly in Eqs. (\ref{4e15}) and (\ref{4e16}).)\par
One is left with the two equations (\ref{4e13c}) and (\ref{4e13d}),
which become simple algebraic equations :
\subequations
\begin{eqnarray}
\label{5e2a}
P_L e^{\displaystyle -2a_3} \phi_0\ -\ M\psi_{B4} &=& 0\ ,\\
\label{5e2b}
P_L e^{\displaystyle -2a_4} \psi_{B4}\ -\ M\phi_0 &=& 0\ .
\end{eqnarray}
\endsubequations
These equations have a nontrivial solution only if $a_3 + a_4$ is a
constant :
\begin{equation} \label{5e3}
a_3 + a_4\ =\ C\ =\ const.\ .
\end{equation}
Then :
\begin{eqnarray}
\label{5e4}
\psi_{B4} &=& \frac {M}{P_L} e^{\displaystyle 2a_4} \phi_0\ =\
\frac {P_L}{M} e^{\displaystyle -2a_3} \phi_0\ ,\\
\label{5e5}
P_L &=& M e^C\ .
\end{eqnarray}
\par
We now check the normalizability of the solution thus found.
For simplicity, we shall consider potentials that are independent of
$P^2$ in the c.m. frame; the corresponding conclusions are not much
affected by an eventual smooth $P^2$ dependence of the potentials.
The normalization condition (\ref{3e12}) becomes :
\begin{eqnarray} \label{5e6}
4\int d^3{\bf x} \bigg [ \ \psi_{B3}^{\dagger} e^{\displaystyle -2a_3}
\psi_{B3} &+& \psi_{B4}^{\dagger} e^{\displaystyle -2a_4} \psi_{B4}\
\bigg ]\ =\ 2P_0\ ,\nonumber \\
4\vert \phi_0\vert ^2 \int d^3{\bf x} \big [\ e^{\displaystyle -2a_3}
&+& e^{\displaystyle 2a_4} e^{\displaystyle -2C}\ \big ]\ =\ 2P_0\ ,
\nonumber \\
8\vert \phi_0\vert ^2 \int d^3{\bf x}\ e^{\displaystyle -2a_3}& & =\
2P_0\ ,
\end{eqnarray}
where Eqs. (\ref{5e1}) and (\ref{5e3})-(\ref{5e5}) were used;
furthermore, the spin 0 projection must be taken in the potentials.
Therefore, $e^{\displaystyle -2a_3}$ must be a rapidly decreasing
function when $\vert {\bf x}\vert \rightarrow \infty\ $, or, equivalently,
$a_3$ must be an increasing function of $\vert {\bf x}\vert $ at
infinity, indicating the confining nature of the potential.\par
To have a more explicit representation of the potentials satisfying
the above conditions, let us consider again the class of potentials
composed of general combinations of scalar, pseudoscalar and vector
potentials [Eqs. (\ref{4e19}) and (\ref{4e20})]. Condition (\ref{5e3})
means that
\begin{equation} \label{5e7}
2(V_1 - V_2)\ =\ C\ .
\end{equation}
Thus, the scalar and timelike vector potentials cannot be chosen
independently from each other.\par
The normalizability condition (\ref{5e6}) implies :
\begin{equation} \label{5e8}
\lim_{\vert {\bf x}\vert \rightarrow \infty } (V_3 - 3U_4 - T_4)\ =\
\infty \ .
\end{equation}
(The spin 0 projection of the potentials has been taken.) This
combination of the pseudoscalar and spacelike vector potentials must
therefore be of the confining type.\par
The component $\psi_{B4}$ [Eqs. (\ref{5e4}) and (\ref{5e5})] of the
wave function then becomes :
\begin{equation} \label{5e9}
\psi_{B4}\ =\ e^{\displaystyle -2(V_3 - 3U_4 - T_4)} \phi_0\ .
\end{equation}
\par
Equation (\ref{5e5}) shows that when the masses of the constituent
particles vanish, then the mass of the bound state also vanishes.
[The fact that the right-hand side of Eq. (\ref{5e6}) vanishes in this
limit should not lead one to the immediate conclusion that this state
disappears from the spectrum. It is its coupling to the axial vector
current which is important on physical grounds, and this coupling involves
the relationship of the wave function to the Bethe-Salpeter wave function
through nonlocal operators, where also $P_L$ is involved \cite{s2}.]\par
As far as the potentials do not have singularities at finite distances,
the function $\psi_{B4}$ [Eqs. (\ref{5e4}) and (\ref{5e9})] does not
vanish at finite
distances and the corresponding wave function $\psi_B$ does not have nodes;
it is then a candidate for the ground state of the spectrum. To conclude
that this is actually the case necessitates a detailed study of the various
potentials in all sectors of quantum numbers. Conditions (\ref{5e8})
and (\ref{5e7}) are not sufficient to guarantee confinement in general.
There are cases of potentials satisfying these conditions, for which
confinement does not occur in a particular sector of quantum numbers or
for which some solutions become unnormalizable. However, there are also
cases for which the above solution is the ground state of the
spectrum; in particular, when the
confining potential is represented by the pseudoscalar potential,
Eq. (\ref{4e18}) can easily be analyzed; in this case all solutions
other than the one found above remain massive in the limit of
vanishing constituent masses.\par
Equations (4.14) also have solutions for which $\psi_{B3} =
\psi_{B4} = 0$ and $\psi_{B1}$ and $\psi_{B2}$ are nonzero. In this case
one finds the solution $P_L = \vert m_1 - m_2\vert
e^{\displaystyle -(a_1 + a_2)}\ $, with $(a_1+a_2)$ equal to a constant.
This solution is, however, unphysical, since it belongs to one of the
unphysical subspaces, where one of the longitudinal momenta, $p_{1L}$
or $p_{2L}\ $, calculated from Eqs. (\ref{2e6}) and (\ref{2e22}), may
become negative \cite{s1}.\par
Finally, the solution found above can also be expressed in the
``canonical'' representation. Taking into account the relationship
(\ref{3e10}), one finds :
\begin{equation} \label{5e10}
\psi \ =\ (1+\gamma_L)\gamma_5 e^{\displaystyle -a_3} \phi_0 \ .
\end{equation}
\par
The massless pseudoscalar bound state solution found in this section
does not of course exhaust all possibly existing solutions. Furthermore,
several types of mechanism may lead to the occurrence of massless
pseudoscalar bound states, in connection with the spontaneous breakdown
of chiral symmetry. Two such mechanisms are : i) the dynamical fermion
mass generation, due to radiative corrections in the fermion self-energy
part \cite{njl}; ii) the fall to the center phenomenon, due to short
distance singularities \cite{mf}. Our solution differs from the above
two in that it is a direct consequence of the particular confining
nature of the interaction and therefore hinges on long distance forces,
rather than on the short distance ones or on the radiative corrections.
The solution corresponding to the pure pseudoscalar interaction case
was studied in detail in Ref. \cite{s3}.\par

\section{Conclusion}

We have shown that, by an appropriate modification of the structure
of the interaction potential, the Breit equation can be incorporated
into a set of two compatible manifestly covariant wave equations,
derived from the general rules of Constraint Theory. The complementary
equation to the covariant Breit type equation determines the evolution
law of the system in the relative time variable and also determines
its relative energy with respect to the other variables. Furthermore,
in this covariant version of the Breit equation, the interaction potential
can be systematically calculated in perturbation theory from Feynman
diagrams by means of a Lippmann-Schwinger-Quasipotential type equation,
relating it to the off-mass shell scattering amplitude.\par
The normalization condition of the Breit wave function indicates
the presence of an interaction dependent kernel in it,
which should be taken into account for consistent evaluations of
physical quantities, like coupling constants, or for the selection of
acceptable (normalizable) solutions to the wave equations.
In this respect, we exhibited, as a straightforward application of the
covariant Breit equation, massless pseudoscalar bound state solutions,
corresponding to a class of confining potentials, essentially
composed of pseudoscalar and spacelike vector potentials with
eventually a particular combination of scalar and timelike vector
potentials.\par
The covariant two-body Breit equation suggests several possibilities
for its generalization to the $N$-body case ($N>2$) or for the
incorporation of external potentials. However, one meets here the
known difficulty of the ``continuum dissolution'' problem \cite{br,sch1},
which prevents the existence of normalizable states. Usually, this
difficulty is circumvented by the introduction of projection operators,
either in the potential \cite{sch2,bb} or in the  kinetic terms \cite{mw}.
It is not yet known whether some local generalization of the Breit
equation may avoid the above difficulty.\par

\end{document}